# Mars in the aftermath of a colossal impact

J. M.Y. Woo[a,b], H. Genda[a,b], R. Brasser[a] and S. J. Mojzsis[c,d]

[a] Earth-Life Science Institute, Tokyo Institute of Technology, Meguro-ku, Tokyo, 152-8550, Japan

[b] Department of Earth and Planetary Sciences, Tokyo Institute of Technology, Meguro-ku, Tokyo 152-8550, Japan

[c] Department of Geological Sciences, University of Colorado, UCB 399, 2200 Colorado Avenue, Boulder, CO 80309-0399, USA

[d] Institute for Geological and Geochemical Research, Research Center for Astronomy and Earth Sciences, Hungarian Academy of Sciences, 45 Budaörsi Street, H-1112 Budapest, Hungary

Abstract

The abundance of highly siderophile elements (HSEs) inferred for Mars' mantle from martian meteorites implies a Late Veneer (LV) mass addition of ~0.8 wt% with broadly chondritic composition. Late accretion to Mars by a differentiated Ceres-sized (~1000 km diameter) object can account for part of the requisite LV mass, and geochronological constraints suggests that this must have occurred no later than ca. 4480 Ma. Here, we analyze the outcome of the hypothetical LV giant impact to Mars with smoothed particle hydrodynamics simulations together with analytical theory. Results show that, in general about 50% of the impactor's metallic core shatters into ~10 m fragments that subsequently fragment into sub-mm metallic hail at re-accretion. This returns a promising delivery of HSEs into martian mantle compared to either a head-on and hit-and-run collision; in both cases, less than 10% of impactor's core materials are fragmented and finally embedded in the martian mantle. Isotopic evidence from martian meteorites, and interpretations from atmospheric mapping data show that a global surface water reservoir could be present during the early Noachian (before ca. 4100 Ma). The millimeter-sized metal hail could thus react with a martian hydrosphere to generate ~3 bars of $H_2$, which is adequate to act as a greenhouse and keep early Mars warm. Yet, we also find that this atmosphere is transient. It typically survives shorter than 3 Myr based on the expected extreme ultraviolet (EUV) flux of the early Sun; if the Sun was a slow rotator an accordingly weaker EUV flux could extend this lifetime to more than 10 Myr. A dense pre-Noachian $CO_2$ atmosphere should lower the escape efficiency of hydrogen by IR emission. A more detailed hydrodynamic atmospheric model of this early hydrogen atmosphere is warranted to examine its effect on pre-Noachian Mars.

1. Introduction

The highly siderophile elements (HSEs), which include the platinum-group metals (Os, Ir, Ru, Rh, Pt, Pd), as well as Au and Re, are strongly depleted in the mantles of Earth, the Moon and Mars relative to chondritic meteorites (Becker et al., 2006; Day et al., 2007, 2016; Tait and Day, 2018). Furthermore, the HSE concentration in both the terrestrial and martian mantles are similar: about 1% of chondritic abundances (Becker et al., 2006; Day et al., 2017; Tait and Day, 2018). During planet formation, metal-silicate partitioning is expected to effectively strip HSEs from silicate mantles into the growing metallic cores. Yet, the observed abundances of HSEs are greatly enhanced relative to their predicted quantities in the silicate mantles of Earth and Mars. Experiments for HSE partition between liquid metal and silicate predict their abundance should be orders of magnitude lower than what is observed (e.g. Kimura et al., 1974; Mann et al., 2012; Rubie et al., 2015). One theory invoked to explain this discrepancy is that the HSEs were delivered after silicate-metal differentiation (i.e. core formation) in the form of a "Late Veneer" (LV) impactor of broadly chondritic composition (Chou, 1978; Frank et al., 2016; Walker, 2009). An alternative mechanism to explain this anomalous enrichment involves inefficient metal-silicate partitioning at high pressures and temperatures (e.g. Murthy, 1991; Righter et al., 2015). A shortcoming of this mechanism, however, is that it cannot currently explain the chondritic relative abundance of the HSEs for both Earth and Mars (Walker, 2009; Bottke et al., 2010; Day et al., 2016; Tait and Day, 2018). Rubie et al. (2016) notes that the partition coefficients of some HSEs at high pressure (Mann et al. 2012) are too low to deplete the HSE abundances enough for the late veneer to overprint them. If core formation were the only HSE depletion mechanism, then we should expect that the HSE abundance pattern would be non-chondritic. This is, however, not the case: it is chondritic, and therefore consistent with a late veneer and not core formation at high pressure. Consequently, there exist two options to interpret these data: either, 1) core formation occurred at low pressure, which is inconsistent with the patterns documented for moderately siderophile elements in mantle samples, or 2) the Hadean Matte hypothesis of O'Neill (1991), which still calls for a late veneer to deliver the HSEs observed in Earth's mantle today.

According to HSE abundances inferred from martian meteorites for Mars' mantle composition, the planet accreted at least 0.5 wt% of material of chondritic composition during the LV stage of late accretion (Walker, 2009; Brandon et al., 2012; Day and Walker, 2015; Tait and Day, 2018). The most likely contribution is computed to be ~0.8 wt% (Day et al., 2016), a value comparable to that estimated for Earth's mass augmentation (~0.3 - 0.8 wt%) from a similar event (Day et al., 2016). In contrast, the estimated LV to the Moon is only 0.02 - 0.035 wt% (Day et al., 2007; Day and Walker, 2015; Kruijer et al., 2015; Touboul et al., 2015). This means that the ratio of accreted mass between Earth and the Moon is 1950 ± 650, which is more than two orders of magnitude higher than the ratio of their gravitational cross sections (~20). As proposed by Bottke et al. (2010), the majority of the mass delivered to Earth after primary accretion could come from a few massive objects the size of Ceres (950 km in diameter) or larger, which lingered behind as left-overs from terrestrial planet formation. The small number of such massive objects in the inner Solar system at that time statistically favors collision with Earth over the Moon (Sleep et al., 1989; Bottke et al., 2010). This hypothesis was further developed by Brasser et al. (2016) who combined the results of N-body and Monte-Carlo impact simulations to conclude that most of the Earth's LV came from a single lunar-sized impactor in the first 100 Myr of the solar system. By adopting the same method for Mars, Brasser and Mojzsis (2017) suggested that a similar but smaller impact affected the red planet. With the size-frequency distribution of the leftover planetesimals assumed to be identical to the current asteroid belt, Mars is expected to have encountered a Ceres-sized object if it accreted 0.7 to 0.8 wt% during the LV (Brasser and Mojzsis, 2017).

If such a colossal impact did occur on Mars, topographical evidence may be imprinted on the crustal surface of Mars. Some studies propose that the martian hemispheric dichotomy exemplified by the northern lowlands region (dubbed the Borealis Basin) was formed by a single giant impact from an object with diameter ranging from 1000 - 2700 km (Wilhelms and Squyres, 1984; Andrews-Hanna et al., 2008; Nimmo et al., 2008; Marinova et al., 2008; Bottke and Andrews-Hanna, 2017; Hyodo et al. 2018), equivalent in size to that estimated by Brasser and Mojzsis (2017). The martian moons Phobos and Deimos may provide further evidence for such a hypothetical colossal impact because the event would likely eject a vast amount of debris into a circum-martian disk to eventually form the martian satellites (Craddock, 2011; Rosenblatt and Charnoz, 2012; Citron et al., 2015; Rosenblatt et al., 2016). Hence, the composition of Phobos and Deimos is predicted to be a mixture of the martian mantle and such an impactor (Rosenblatt et al., 2016; Hyodo et al., 2017). The future sample return mission of JAXA's MMX (Martian Moons eXploration) aims in part to understand the physical and chemical nature of these moons, and thus provide insight on whether a colossal impact may be responsible for their origin.

The minimum timing for this postulated LV impact on Mars can be potentially constrained from ancient igneous zircons in martian meteorites. Zircon U-Pb geochronology performed on martian meteorite NWA7533 yields 4428 Ma ages (Humayun et al., 2013). Recent results reveal even older ages, ranging from 4430 – 4480 Ma in the meteorite NWA7034 (Bouvier et al., 2018), which is a paired with NWA7533 (Agee et al., 2013). Zircons in these meteorites are present in matrix material as fragments as well as in evolved igneous clasts perhaps formed by re-melting of the primary martian crust either at depth in the presence of volatiles (Stolper et al., 2013), or by differentiation of large impact melt sheets (Humayun et al., 2013). Indications are that a crust has existed on Mars since 4547 Ma, which implies that magma ocean crystallization of Mars occurred rapidly after Solar System formation: the primordial martian crust formed within 20 Myr (Bouvier et al., 2018). Although this rapid magma ocean crystallization is fully consistent with theoretical models (e.g. Elkins-Tanton, 2008; Lebrun et al., 2013; Hamano et al., 2013), it appears at odds with the Sm-Nd systematics of the martian SNC meteorites, which require at least a shallow long-lived magma ocean for ~60-100 Myr (Debaille et al., 2007; Borg et al., 2016). Yet, the latter conclusion is based on the abundance of short-lived radionuclides in young martian meteorites and therefore may insinuate a younger fractionation event in the regional mantle scale rather than the actual martian crust solidification timescale (Bouvier et al., 2018). The relatively late formation of the zircons has been attributed to a LV colossal impact near 4480 Ma that melted a part of the martian crust (Bouvier et al., 2018).

A Ceres-sized impactor that collided during the pre-Noachian eon (4500-4100 Ma) should profoundly affect the early surface environment of Mars. Objects with diameter >1000 km are expected to be differentiated (Stevenson, 1981), having a metal core and a silicate mantle. Genda et al. (2017a) showed that under certain conditions a dense (~80-90 bar) hydrogen atmosphere can be generated after a differentiated lunar-sized object collided with Earth during a late accretion. Depending on Solar extreme ultraviolet (EUV) models, this post-impact atmosphere may have persisted on Earth for ~100-200 Myr. The impactor's iron core materials are flung into geocentric orbit upon impact, and then sheared and fragmented into smaller drops, which later descended back to the terrestrial surface as mm-scale metallic hail. Hydrogen gas is generated through oxidation of falling metal to the Earth either by the reaction with a global surface ocean or an oxidized Hadean mantle (Abe, 1993; Genda and Abe, 2005).

Provided that pre-Noachian Mars had adequate global surface water (or surface ice) at time of an LV colossal impact, a similar, but less massive collision (Brasser and Mojzsis, 2017) ought to likewise yield a transient hydrogen-rich atmosphere. Hydrogen molecules collide frequently under high pressure (~2 bar) and hence possess transient dipole moments that induce infrared absorption (Pierrehumbert and Gaidos, 2011). A hydrogen greenhouse on early Mars (Ramirez, 2017; Ramirez and Kaltenegger, 2017) could in principle provide an atmospheric factory for organic chemistry and maintain an early warm climate, even if it was for a short time. This has obvious implications for the origin of life on Mars in the planet's first hundred million years.

Here, we analyze the fate of a mechanically-disrupted iron core from a leftover Ceres-sized planetary embryo striking Mars during the pre-Noachian. Our study employs smoothed particle hydrodynamic (SPH) impact simulations as well as analytical estimations of the post-collision evolution of the impactor's core materials with a postulated hydrosphere on pre-Noachian Mars. We conclude with recommendations for future studies to test predictions arising from this model.

2. Methods

The smoothed particle hydrodynamics (SPH) method (e.g. Monaghan, 1992), is a flexible Lagrangian approach for solving hydrodynamic equations. Our numerical code is identical to that used in Genda et al. (2015, 2017a). Our code calculates a purely hydrodynamic flow with self-gravity, but without material strength. For brevity, we refer the reader to these published studies for details on the code, its methods and tests. Here we briefly describe the initial conditions of our SPH simulations.

In our analysis, we consider a ~1000 km diameter impactor striking Mars under different impact angles, $\theta$, and impact velocities, $v_{imp}$. We propose that this object be named *Nerio*. She was Mars' consort, an ancient war goddess, and the personification of courage.

She was the partner of Mars in ancient cult practices. Relative to the horizontal reference frame, 8 simulations were performed with $\theta = 0°$ (a head-on collision), 30°, 35°, 40°, 45° (statistically most-likely; Shoemaker, 1962), 50°, 55° and 60°. We set $v_{imp} = 10$ km/s (~$2v_{esc}$, where $v_{esc} \sim 5$ km/s is the surface escape velocity of Mars) which is the mean value obtained from N-body simulations of leftover planetesimals from terrestrial planet formation (Brasser et al., 2016). We also perform three additional simulations with $v_{imp} = 7$ km/s (~1.4 $v_{esc}$), 13 km/s (~2.6 $v_{esc}$) and 16 km/s (~3.2 $v_{esc}$), but restrict these to $\theta = 45°$. The mass of the target (Mars) is set to 6.4 x $10^{23}$ kg and the impactor is 0.3 wt% of Mars (2.0 x $10^{20}$ kg). The number of SPH particles used for the target and impact are 3 million and 9600 particles, respectively. Both objects are assumed to be completely differentiated: they have a chondritic initial composition with a 30 wt% iron core and a 70 wt% silicate mantle.

Based on the relative abundances of the HSEs as measured in terrestrial mantle-derived rocks, Earth's LV augmentation is suggested to consist of mainly enstatite chondritic material (Fischer-Gödde and Kleine, 2017). Current isotopic studies coupled with dynamical analysis performed to study the origin of Mars provide no meaningful limitation on the composition of its LV impactor (Brasser et al., 2018). Hence, we follow the approach of Genda et al. (2017a) and consider a bulk composition of enstatite chondrite with 30 wt% of its mass as reduced iron sequestered in its core (Wasson and Kallemeyn, 1988). We ignore the spin effect for pre-impact objects. The surface velocity of a *Nerio*-scale spinning impactor and target Mars is much less than the impact velocity (Genda et al., 2017a; Canup, 2008). The initial internal energy is assumed to be 1 x $10^5$ J/kg (Genda et al., 2017a), and we calculate vibrations of the impactor and target until the particle velocity becomes slower than 100 m/s; this value is much lower than the impact velocity.

We subsequently use these relaxed bodies for impact simulation. The calculated diameter of the gravitationally relaxed impactor is 1040 km. We performed the impact simulations over a period of $10^5$ sec, or about 28 hours.

3. Results and Discussion

*3.1 Collision Outcome*

Snapshots of our SPH simulations for a collision between a *Nerio*-sized impactor and Mars are shown in Fig. 1. The collision results are broadly similar to those reported in Genda et al. (2017a), where a colossal impact was considered between Earth and a 3000 km diameter object with 0.01 Earth's mass. Figure 1(a) shows the time series of snapshots for the case of $v_{imp}$ = 10 km/s (~2 $v_{esc}$) and $\theta = 45°$. After the initial collision with Mars, the impactor undergoes elongation due to the mechanical tearing of both its core and mantle; a significant amount of the impactor's material is ejected onto martian orbit from the collision site. A minority fraction of the ejected materials gain enough kinetic energy during the collision to escape from Mars' gravity. The majority of the ejecta re-accrete back to Mars and blanket its entire surface, with the highest portion being re-accreted onto the surface close to the collision site. A portion of the ejected impactor's materials are shocked after impact and then further fragment into smaller pieces from pressure release. Iron core material of the impactor experiencing fragmentation is shown as yellow particles in Fig. 1. This material pollutes the mantle of Mars with HSEs by re-accreting back onto the whole surface of Mars. Fig. 2 depicts the number of fragmented impactor's iron particles that are finally embedded in the martian mantle after 5.58 hr as a function of longitude for an impact angle of θ = 45º and $v_{imp}$ = 10 km/s (~2 $v_{esc}$). We note that the distribution of fragments are global but non-uniform, which concentrates near the expected impact side (from -50º to 50º from the negative y-axis). Our results are similar to Marchi et al. (2018), which shows that material from large planetesimals (≥ 1500 km) concentrates within localized domains of Earth's mantle after colliding with Earth during post-Moon formation period. This could explain the observed $\varepsilon^{182}W$ isotopic anomalies in terrestrial rocks due to incomplete mixing (Willbold et al., 2015). On billion-year timescales the distribution of HSEs could possibly be homogenized by martian mantle convection.

Snapshots of our impact simulations with other values of $\theta$ and $v_{imp}$ are given in Fig. 1(b). The collision outcome strongly resembles Genda et al. (2017a) for the case of the Earth's collision with its LV impactor *Moneta* and so we limit the results to 5.58 hr from the start of the simulations. In the $\theta = 0º$ scenario (head-on collision), the core of the impactor merges with the martian core rapidly and nearly all of the impactor's mantle material is confined – initially – to one hemisphere of Mars' mantle. Fragmentation of the impactor's iron core material is rare in head-on collisions since no SPH particles from that region are ejected to experience pressure release. The efficiency of HSEs delivery is very low because almost none of the impactor's core material remains in the martian mantle. Collision outcomes of $\theta = 30º$, 35º and 40º are similar to one another. We find that most of the impactor's mantle (>70%) and all core material is gravitationally bound to Mars when $\theta < 40º$. Compared to the nominal $\theta = 45º$ case, we observe that far fewer tracer particles for the fragmented impactor's core (yellow) in our SPH code exist in the martian mantle. Therefore, the delivery of HSEs is not efficient when $\theta \leq 40º$. More fragmented impactor core particles (yellow) exist in the martian mantle when the $\theta = 50°$ compared to the other lower $\theta$ cases. The *total* number of the bound impactor's SPH particles (the total number of blue, red and yellow SPH particles in the martian mantle) is, however, *lower* in the $\theta = 50°$ scenario. An increase of $\theta$ up to $55°$ and $60°$ yields even fewer bound impactor's materials. This is because the collision becomes nearly hit-and-run and therefore the impactor only grazes the martian surface when $\theta$ is high. Hence a higher portion of the impactor's materials escapes from Mars permanently. Therefore, when $\theta > 50°$ the collision is basically a hit-and-run with impactor's core and mantle strewn beyond Mars into the mid-Solar system.

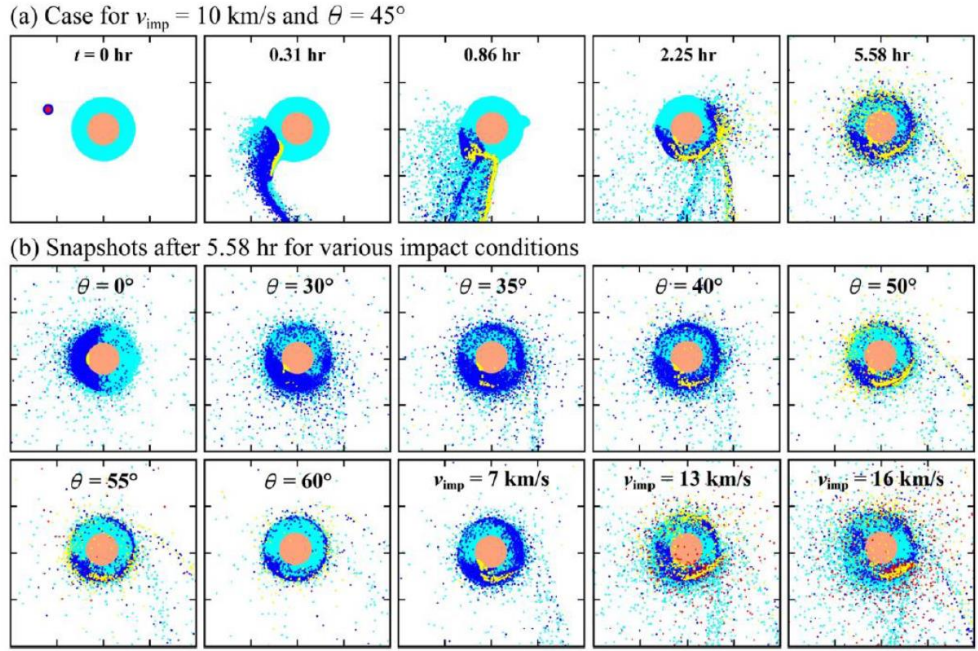

Figure 1: Snapshots for a collision of a *Nerio*-scale impactor (0.3% of Mars' mass) onto early Mars. (a) The time series of snapshots for the case of the impact velocity, $v_{imp}$ = 10 km/s (~$2v_{esc}$, where $v_{esc}$ is the surface escape velocity of Mars) and the impact angle, $\theta$ = 45°. (b) Snapshots after 5.58 hr for various $\theta$ ranging from 0° to 60° with $v_{imp}$ = 10 km/s, and various $v_{imp}$ ranging from 7 km/s to 16 km/s with $\theta$ = 45°. Mantle and core materials for the impactor are colored blue and red, respectively, and those for Mars are light blue and orange. Impactor's core materials that have experienced fragmentation are colored yellow (See the text in Section 3.1 for details). The snapshots are NOT cross sections, but all SPH particles are layered on top of one another in the order of martian mantle, impactor's mantle, martian core, impactor's core (red and then yellow), so that iron particles are clearly seen. The interval of tics in each snapshot is 5000 km or 1.47 martian radii.

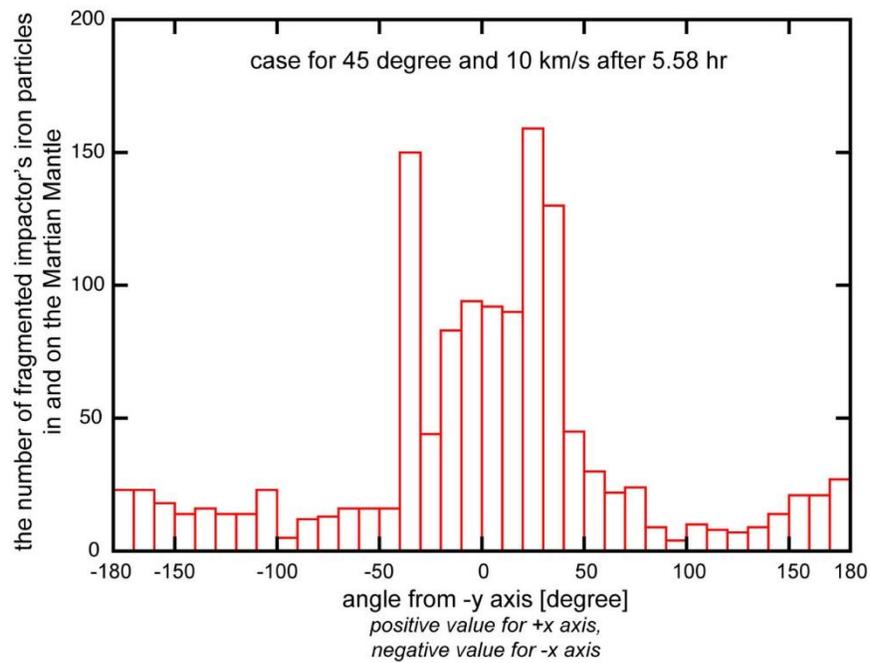

Figure 2: The number of fragmented impactor's iron particles in and on the martian mantle at different longitude from the negative y axis for impact angle, $\theta$= 45° and impact velocity, $v_{imp}$ = 10 km/s (~2 $v_{esc}$) after 5.58 hr. The distribution of fragments are global but concentrated from -50° to 50°.

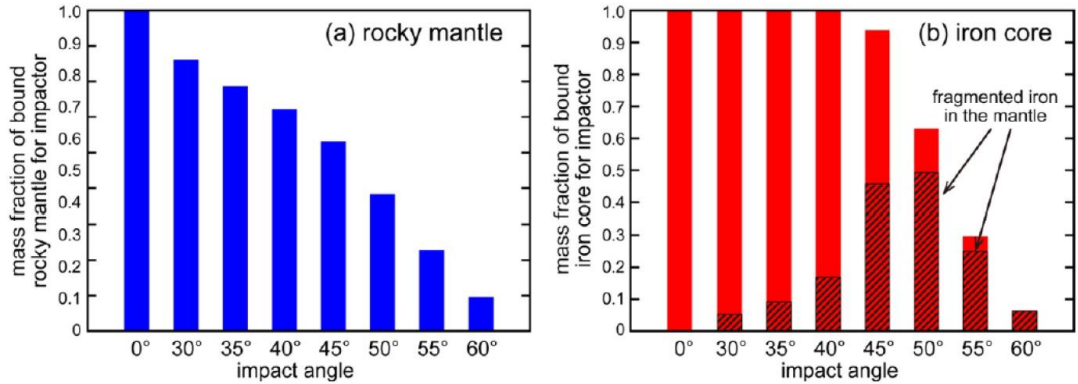

Figure 3: Mass fraction of impactor's materials that are gravitationally bound to Mars after collisions as a function of impact angle, $\theta$. The impact velocity, $v_{imp}$, is the same for all $\theta$ (10 km/s, ~2 $v_{esc}$). The left panel pertains to the impactor's silicate mantle material, while the right panel is for the iron core. Some of the impactor's core materials that are gravitationally bound to Mars experience pressure release after the shock, which lead to fragmentation (See the text in Section 3.1 and 3.2 for details). Shaded region represents the mass fraction of fragmented impactor's iron materials that do not merge into the martian core, but are available for suspension in the martian mantle.

The relation between the amount of impactor's materials bound to Mars and $\theta$ is provided in Fig. 3. The left panel depicts the mass fraction of the impactor's rocky mantle that became bound to Mars after collisions as a function of $\theta$; the right panel depicts the same for the impactor's iron core. The trend of decreasing mass fraction of bound impactor's materials with increasing $\theta$ is clear. Moreover, for the same $\theta$, the mass fraction of the bound impactor's iron core is usually higher than that of the bound impactor's mantle. Since the iron core is enveloped by the mantle, it is therefore more difficult for the iron core to escape Mars' gravity (e.g. Genda et al., 2017a). The shaded regions in the right panel of Fig. 3 represent iron core material of the impactor that experiences fragmentation and finally embedded in the martian mantle; this material is represented by the yellow particles staying in the martian mantle or floating near the martian surface in Fig. 1. The shaded region increases with $\theta$ because a higher fraction of shocked iron core materials are ejected and then undergo pressure release when $\theta$ is larger. The maximum mass of bound fragments is reached at 50°. The shaded region subsequently decreases with $\theta$ from $\theta = 50°$ to 60° because most of the fragmented iron core material of the impactor escapes from the system instead of becoming embedded in the martian mantle. We do not discuss the fate of the non-fragmented impactor's iron because at most ~3% of the impactor's core particles stay in the martian mantle as non-fragmented iron for each $\theta$ case. Hence non-fragmented iron would not contribute much to the delivery of HSEs into the martian mantle and the formation of a hydrogen atmosphere (see section 3.3).

We also perform additional simulations with different $v_{imp}$ (but only for $\theta = 45°$); their snapshots are shown in Fig. 1(b). The total number of the impactor's SPH particles (blue, red and yellow) that are embedded in the martian mantle decreases with increasing $v_{imp}$. Fig. 4 shows this further when we investigate the mass fraction of bound materials for the impactors' mantle and iron core as a function of $v_{imp}$. Results in the left panel are explained by the impactor's materials having a higher kinetic energy when $v_{imp}$ is higher and are thus more likely to escape Mars' gravity. We also found that the fragmented impactor core (yellow particles) exist with the highest abundance when $v_{imp} = 10$ km/s and 13 km/s: nearly half of the fragmented impactor's iron core finally ends up suspended in the martian mantle when $v_{imp} = 10$ km/s and 13 km/s (Fig. 4). This fraction drops to < 30% when $v_{imp} = 7$ or 16 km/s. In the case of lower $v_{imp}$, fewer impactor's iron core materials are ejected into the open space after collision because of their lower kinetic energy and therefore limiting the fragmentation of the iron materials. Higher $v_{imp}$ leads to fragmentation of a larger percentage of the impactor's iron core. In the case of $v_{imp} = 16$ km/s, however, most of the fragmented iron core of the impactor escapes from the system due to the high kinetic energy of the impact and Mars' low surface gravity. These escaped particles are consequently ignored in the shaded region of Fig. 4(b).

In summary, in the impact geometry that is statistically most likely ($\theta = 40$-$50°$ and $v_{imp} = 10$-$13$ km/s), our simulations inject a substantial amount of fragmented iron (~50% of the impactor's iron core) into the mantle of Mars. The efficiency of HSEs delivery to martian mantle dramatically decreases when $\theta \leq 40°$ since less than 20% of the impactor's iron core materials are fragmented and finally suspended in the martian mantle. Similar to low $\theta$ case, high $\theta$ cases ($\theta \geq 55°$) also do not favor HSEs delivery since most of the impactor's core materials escape from the system after the giant impact.

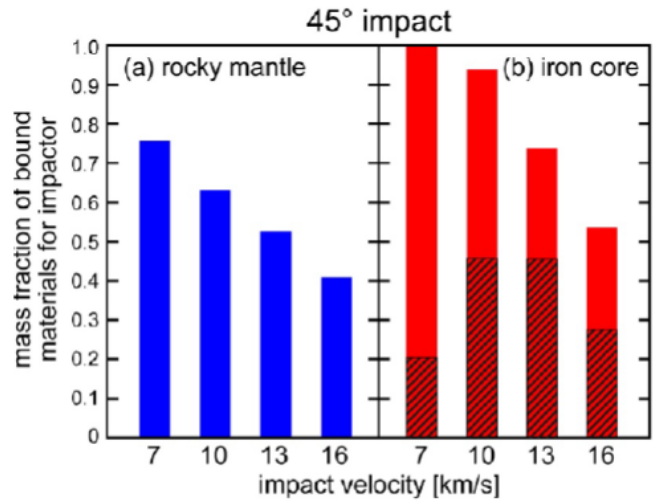

Figure 4: Same as Fig. 3, but the dependence of the impact velocity, $v_\text{imp}$, is shown. The impact angle, $\theta$, is fixed to be 45º.

*3.2 Size of impactor's iron fragments after the giant impact*

In the common oblique collision ($\theta \sim 45°$) scenario, the shock wave generated during the collision passes through the iron core and hence creates a high peak pressure (~ 100GPa) within it; this high peak pressure is below the pressure for shock-induced incipient melting (~220 GPa) and complete melting (~260 GPa) for cold iron (Melosh, 1989), however. We note that some studies propose that small cores such would exist for a *Nerio*-like object can remain molten up to ~100 Myr after CAI formation (e.g., Neumann et al., 2014). At the very least, the impactor's core should be in a hot solid state, in which case re-melting of the iron core by shock passage (~ 100 GPa) is expected. Hence, we assume here that the impactor's core is liquid just after the impact. The molten iron core can thus be pulled apart into an elongated shape and ejected from the near-surface region of target Mars under the dynamical physical conditions we find in our model. After ejection into open space, the iron core material expands due to its high residual internal pressure. This state induces shear stress on the molten core material and result in its fragmentation.

Due to the computationally intensive nature of the high resolution simulations required to directly analyze the fragmentation process, we instead traced the time-dependent evolution of pressure for all SPH particles in our simulations. The expansion of core material from residual internal pressure drops from shock-induced pressure (~100 GPa) to extremely low pressure (~0 GPa). We assume that fragmentation occurs when the pressure of a shocked SPH particle decreases down to 0.1 MPa (1 bar) during the pressure release process (see Genda et al., 2017a). Our tracked SPH particles that suffer pressure release are shown as yellow in Fig. 1; their mass fractions that finally embedded in martian mantle are indicated by shaded regions in Fig. 3(b) and 4(b) (see the previous subsection).

The size of iron fragments, however, cannot be directly computed from the SPH simulations due to limitations of our model resolution and computational feasibility. The dimensions of a single SPH particle is expected to far exceed that of the iron fragments. We thus estimate the fragment size by considering the balance between the surface tension of liquid iron, σ, and the local kinetic energy ($1/2\ mv^2$) induced by the local shear velocity $v = \dot{\varepsilon}\, d$, where $\dot{\varepsilon}$ is the strain rate of the expanding molten iron blob and $d$ is the typical size of molten droplets at fragmentation. The size of the molten iron droplets ejected by an impact is

$$d = \left(\frac{40\sigma}{\rho \dot{\varepsilon}^2}\right)^{1/3} \tag{1}$$

(Melosh and Vickery, 1991), where $\rho = 7000$ kg/m$^3$ is the density of the iron droplets and σ = 2 N/m for liquid iron (Keene, 1993). The $\dot{\varepsilon}$ tensor for each SPH particle can be computed in the SPH code by the method described in Section 2.3 of Genda et al. (2017a).

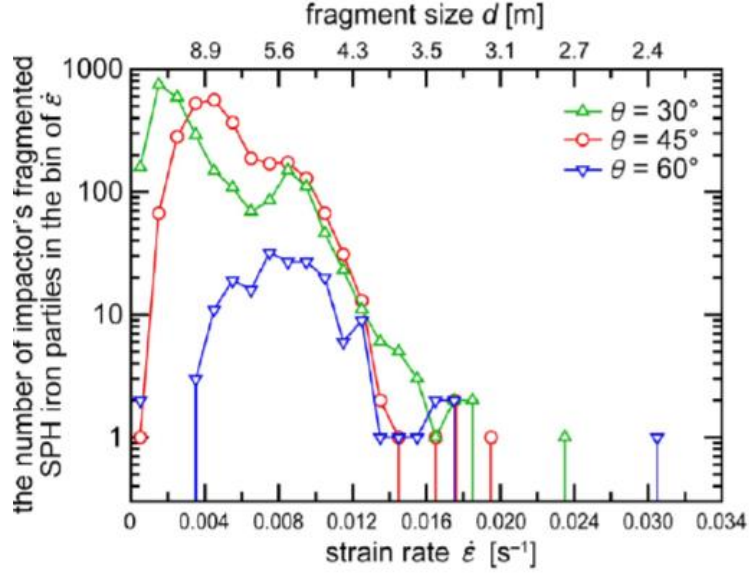

Figure 5: Distribution of the strain rate, $\dot{\varepsilon}$, for the impactor's core material at the time of fragmentation. The collisions for the impact angle, $\theta = 30°$, $45°$ and $60°$ with the impact velocity, $v_{imp} = 10$ km/s are plotted. The bin size of $\dot{\varepsilon}$ is 0.001. The fragment size, $d$, is estimated by the balance between the surface tension of the liquid iron, $\sigma$, and $\dot{\varepsilon}$ (see Eq. (1)).

Fig. 5 shows the distribution of the strain rate, $\dot{\varepsilon}$, for the impactor's core material at the time of fragmentation. The calculated strain rate is ~0.001 - 0.03 s$^{-1}$, which is consistent with the value estimated by the following simple physical consideration: $\dot{\varepsilon} \sim v_{imp}/D_{imp} = 0.01$ s$^{-1}$ (Melosh and Vickery, 1991), where $v_{imp} = 10$ km/s and $D_{imp} = 1040$ km, which is the impactor's diameter in our simulation. The size of the iron fragments can be estimated from Eq. (1), which are shown in Fig. 5. The peak of the green curve indicates that most of the fragments have a size of ~14 m when $\theta = 30°$. Increasing $\theta$ to $45°$ and $60°$ results in an even smaller fragment size. The peak of the red curve ($\theta = 45°$) and the blue curve ($\theta = 60°$) are ~10 m and ~6 m, respectively. Since the impactor suffers a greater extent of elongation with a larger $\theta$ collision, the iron core material of the impactor experiences a faster change in strain. Although with larger $\theta$ the mass fraction of the impactor's iron core that re-accretes onto Mars is lower, the fragment size is generally smaller ("metallic hail"). Taking the results of the statistically most likely case ($\theta = 45°$), we conclude that the typical fragment size of the impactor's iron core is about 10 m after the giant impact, which is similar to the fragment size from the disrupted core of a lunar-sized impactor during its collision with Earth (Genda et al., 2017a). We need to emphasize that these 10 m iron fragments are not their final size when they settle on Mars' surface because they experience further fragmentation during the re-accretion process. If iron fragments are ejected from the forming crater during re-accretion, the typical size of ejected fragments can also be estimated from Eq. (1). Given a strain rate $\dot{\varepsilon} \sim v_{imp}/D_{imp} = 252$ s$^{-1}$, where now $v_{imp} \sim 2.52$ km/s, which is half of Mars' escape velocity, and $D_{imp} = 10$ m is the typical fragment size: Substituting $\dot{\varepsilon} = 252$ s$^{-1}$ into Eq. (1), we obtain d ~ 6 mm. Therefore, we predict that a mm-sized metallic hail returns to Mars after further fragmentation.

*3.3 A temporary hydrogen atmosphere as a byproduct of colossal impact*

If sufficient mm-sized iron fragments rain back onto the surface of Mars a transient hydrogen atmosphere can potentially be generated by the reduction of water:

$$Fe + H_2O \rightarrow FeO + H_2 \qquad (2)$$

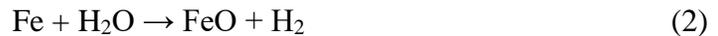

This reaction should take place mostly between the mm-sized metallic hail and postulated global surface hydrosphere (or cryosphere) on pre-Noachian Mars, provided that the water of the terrestrial planets was predominantly delivered during the main accretion stage of the planets as opposed to the late veneer (e.g. Greenwood et al., 2018). According to isotopic studies (Sanloup et al., 1999; Dauphas and Pourmand, 2011; Brasser et al., 2018) together with dynamical simulations (Woo et al., 2018), Mars accreted a higher portion of ordinary chondrite-like material (~50%) than Earth (< 25%) as its building blocks. In contrast, Earth accreted mostly from enstatite chondrite-like sources (Dauphas, 2017; Brasser et al., 2018; Javoy et al., 2010), which has a far lower water content (~0.01 wt%) (Hutson and Ruzicka, 2000) than ordinary chondrite (~0.1 wt%) (McNaughton et al., 1981; Robert et al., 1977, 1979). Based on these analyses, we conclude that the higher portion of ordinary chondrite (Woo et al., 2018; Brasser et al., 2018) together with the low fraction of water-rich carbonaceous chondrite (5 – 10 wt%) (Kerridge, 1985; Robert and Epstein, 1982) delivered most of the martian water. The total estimated initial martian water budget from the coupled dynamical-cosmochemical modeling results of Brasser et al. (2018) and Woo et al. (2018) corresponds to ~600 ppm, which is somewhat higher than Mars' current computed bulk water content of 300 ±150 ppm (Taylor, 2013). If most of the water in the martian mantle outgassed during magma ocean solidification and condensed within ~0.1 Myr after the magma ocean solidification (Lebrun et al., 2013) at ca. 4547 Ma (Bouvier et al., 2018) then it possibly created a global surface water reservoir on very early Mars equivalent to a ~3 km thick ocean (or thicker ice sheet). Mapping of atmospheric water ($H_2O$) and its deuterated form (HDO) shows a high D/H ratio in Mars' atmosphere, which indicates Mars experienced substantial water loss since the pre-Noachian and hence an ancient water reservoir covering 20% of the martian surface at 4500 Ma (Villanueva et al., 2015). This is also supported by the fractionated xenon isotopes exist in ancient martian meteorites, which indicates intense water loss within the first few million years of Mars' formation (Cassata, 2017). Plausibility arguments have been made for global surface water to have existed on Mars during the pre-Noachian (see Ramirez and Craddock, 2018 for a recent perspective).

Assuming that all fragmented iron that finally ended up suspended in the martian mantle reacts with the postulated surface water reservoir, we calculate that ~3 bars of hydrogen can be created by reducing ~$10^{20}$ kg of water (~150 ppm) from ~3 x $10^{20}$ kg of iron material from the impactor's core. Such a pre-Noachian hydrogen atmosphere is dense enough to act as a greenhouse gas and temporarily increase the surface temperature of a planet above the freezing point of water (Pierrehumbert and Gaidos, 2011). We also find that this process yields $10^{21}$ kg FeO, which accounts for ~0.14 wt% of the martian mantle's FeO content (~18 wt% of the martian mantle) measured by Mars' orbiter and rovers, as well as that measured from martian meteorites (Taylor, 2013, and references therein). Thus far, there are no obvious physical or chemical reasons to preclude the formation of such an atmosphere on very early Mars, except that such an atmosphere should have been short-lived.

The more intense extreme ultraviolet (EUV) of the young sun (Ribas et al., 2010) leads to the rapid decline of the hydrogen atmosphere through the process of hydrodynamic escape (Hamano et al., 2013; Genda et al., 2017b). The escape flux of hydrogen can be estimated by,

$$\phi_{H_2} = \frac{\varepsilon_{eff} f_{EUV}(t) R}{4GMm_{H_2}} [m^{-2}s^{-1}] \quad (3)$$

(Watson et al., 1981), where $\phi_{H2}$ is the escape flux of $H_2$, $G$ is the gravitational constant, $R$ is the planetary radius, $M$ is the planetary mass and $m_{H_2}$ is the molecular mass of $H_2$, $f_{EUV}(t)$ is the EUV energy flux received by Mars and $\varepsilon_{eff}$ is the escape efficiency, which represents how large a proportion of the received EUV energy is available for escaping $H_2$ molecules. We adopt $\varepsilon_{eff} = 0.3$ (Sekiya et al., 1980; Genda et al., 2017b) and use the scaling laws of the Solar EUV energy flux of Ribas et al. (2005), which is

$$f_{EUV}(t) = 0.03 \left(\frac{t}{1Gyr}\right)^{-1} \quad (4)$$

where $t$ is the age of the Sun. Since the original $f_{EUV}$ in Ribas et al. (2005) is expressed for the flux received by Earth, we include a factor $(1AU/1.5AU)^2$ in Eq. (4) to account for the flux received by Mars.

As mentioned in Section 1, the LV impact on Mars was likely to have occurred no later than 4.48 Ga (Bouvier et al., 2018), which is ~80 Myr after the Sun's birth. Thus, we adopt $t = 0.08$ Gyr as the starting time of the giant impact. We should mention that the $f_{EUV}$ scaling law of Ribas et al. (2005) was derived according to Solar like stars that are at least 100 Myr old. The time of the hypothetical LV giant impact occurred when the Sun was only 80 Myr old, which is not within the age range considered in Ribas et al. (2005). Nonetheless, the Sun was likely to be in the main sequence stage of its evolution some 50 Myr after its formation (e.g. Siess et al., 2000). Extrapolating the scaling law to $t < 100$ Myr matches the predicted EUV flux of the Sun in between 50 to 100 Myr after birth (Pizzolato et al., 2013). Therefore, it is valid to adopt the scaling law in Eq. (4) when computing the life time of the hydrogen atmosphere. Integrating Eq. (3) with the time profile of (4) provides us with the finishing time of the hydrodynamic escape and hence a postulated lifetime of the hydrogen atmosphere. From this we estimate the hydrogen atmosphere persists no more than 3 Myr, which nearly is two orders of magnitude shorter than what has been proposed via core disruption of a lunar-sized impactor in collision with Earth (~100 Myr; Genda et al., 2017a). The major reason for this discrepancy is that the mass of the iron core of a *Nerio*-scale impactor is a mere ~3% of that of a lunar-sized impactor and a mere 0.3 wt% of Mars' mass compared to the ~1 wt% of Moneta compared to Earth (Genda et al., 2017b). As a result, the mass of the hydrogen atmosphere, and thus its lifetime, is directly proportional to the total mass of the impactor's iron core reacting with the global surface water on each planet.

Several assumptions are at play that can profoundly affect the lifetime estimates of this impact-generated hydrogen atmosphere. The most important of these is the assumed Solar EUV flux. We make use of the EUV flux scaling law deduced by Ribas et al. (2005). They only examine seven main sequence G-type stars (including the Sun) to compute the scaling laws of stellar fluxes at different wavelengths. Yet, stellar rotational evolution is one of the principal factors that determine its radiative output (Johnstone et al., 2015; Tu et al., 2015). Stars in young clusters (<500 Myr) have a wide range of rotational rates (Soderblom et al., 1993). According to the EUV flux evolution model of Tu et al. (2015), the EUV flux evolution of stars with different initial rotation periods diverge between the ages of 30 to 1000 Myr. The Tu et al. model takes a broad observation sample into account:

they use data from eight young stellar clusters with ages from 30 Myr to 620 Myr. A Solar mass star with a fast initial rotation rate (initial period <1 day) generates ~10 times more EUV flux than a slow rotator (initial period comparable to the current Sun's). About 70% of the Solar mass stars studied by Johnstone et al. (2015) belong to this class of slow rotators. Therefore, if the young Sun had a low spin rate initially, its EUV flux intensity would only be about 20 times higher than its current value, instead of about a hundred times higher according to the scaling laws of Ribas et al. (2005). The corresponding survival timescale of a hydrogen atmosphere on Mars could thus be about five times longer (~10 to 12 Myr) under a weaker EUV flux.

Another factor affecting the lifetime of the hydrogen atmosphere is the abundance of $CO_2$ in the martian atmosphere prior the LV impact. Our calculation with Eq. (3) and (4) does not include other gases that may exist before this event. While assuming that a global surface water exists in liquid form on Mars during the first tens of millions of years of Mars' history, keeping the average global surface temperature of Mars above the freezing point requires about 1 - 5 bars of $CO_2$ (Kasting, 1991; Forget and Pierrehumbert, 1997). If sufficient $CO_2$ existed before the Late Veneer impact, its 15 µm band infrared emission is the major cooling process in the lower thermosphere of Mars (Gordiets et al., 1982; Lammer et al., 2006) and hence the escape efficiency, $\varepsilon_{eff}$ in Eq. (3), decreases. The solidification of the magma ocean outgasses volatiles and cumulatively yields as much as ~100 bars of atmosphere composed of $H_2O$ and $CO_2$ (Elkins-Tanton, 2008). According to the hydrodynamical model of Erkaev et al. (2014) retaining this thick initial atmosphere is problematic because it can fully escape within 10 Myr due to the strong EUV of the young

Sun; not to mention that ~100 bars of volatile in ancient martian atmosphere is likely an over-estimate. Impact erosion during the first ten million years of Mars' formation (e.g. Melosh and Vickery, 1989; Schlichting et al., 2015) is another plausible factor mitigating an initial thick $CO_2$ atmosphere in the pre-Noachian. Regardless, there is abundant evidence of volcanic activity on Mars that has the potential to re-supply the escaped $CO_2$ and help offset the loss of a thick atmosphere over geologic time (Michalski et al., 2018, and references therein). It is still unclear precisely when the volcanic activity of Mars began. The Tharsis bulge, which is the largest volcanic complex on Mars, is thought to have initiated its formation during the Noachian epoch (Anderson et al., 2001; Werner, 2009). Similar tectonic evidence is lacking for the pre-Noachian epoch because of the postulated LV giant impact or other bombardment events. Colossal impact(s) possibly triggered crustal re-melting and therefore may have erased the corresponding earliest volcanic features on Mars.

4. Conclusions

We performed SPH simulations to study the consequence of a late veneer colossal impact between Mars and *Nerio*, a ~1000 km diameter impactor that delivered part of the martian Late Veneer. We found that about half of the impactor's core materials undergo fragmentation and become bound to Mars' mantle after impact in the collision scenario that is statistically most likely ($v_{imp}$ = 10 km/s and $\theta$ = 45º). These iron fragments are able to enrich the martian mantle with highly siderophile elements. The typical size of one of the fragments is 10 m after the first collision. These fragments are further torn into mm-sized iron blobs during re-accretion onto Mars. Assuming that global surface water existed in either ice or liquid form during the late veneer, the re-accreted mm-sized iron blobs reacted with the surface water reservoir and created ~3 bars of hydrogen, which is thick enough to act as a greenhouse gas. This hydrogen atmosphere is expected to have fully escaped within 3 Myr, which makes it difficult for it to have had a significant effect on the surface environment of Mars for a long period. However, making assumptions with a slower rotating young Sun with weaker Solar EUV radiation and an initial thick carbon dioxide atmosphere would significantly extend the lifetime of the hydrogen atmosphere to more than 10 Myr. Given the greenhouse nature of hydrogen gas and its implication for biopoesis on early Hadean Earth (e.g. Urey, 1952; Miller, 1953), we call for further study on the possible generation of an early hydrogen atmosphere and its effect on pre-Noachian Mars through more detailed hydrodynamical atmospheric models with different assumptions of early solar EUV radiation and initial atmospheric composition of Mars. More accurate measurements on the abundances and isotopic anomalies of HSEs and moderately siderophile elements (e.g. $^{92}$Mo and $^{64}$Ni) from martian meteorites is warranted in order to understand the martian late veneer delivery process.

Acknowledgments

We thank Shigeru Ida for his helpful advice on our numerical simulations, Oleg Abramov on discussions of martian impact bombardment and Ramses Ramirez on discussions of hydrodynamic escape. Comments to the manuscript by Jay Melosh, James Day, Fred Moynier and two anonymous reviewers helped improve the work. This study was supported by Grant-in-Aids for Scientific Research from the Japan Society for Promotion of Science (JP15K13562, JP 17H02990, JP 17H06457) to HG and (JP 17KK0089) to RB. SJM is also indebted to the Earth-Life Science Institute (ELSI) at the Tokyo Institute of Technology for sabbatical support in 2015-2016 during which time this project was realized. SJM further acknowledges support from the NASA Solar System Workings program (NNH16ZDA001N-SSW) for our on-going investigations of terrestrial-type planetary bombardments. SJM and RB acknowledge the Collaborative for Research in Origins (CRiO), which is supported by The John Templeton Foundation (principal investigator: Steven Benner/FfAME): the opinions expressed in this publication are those of the authors, and do not necessarily reflect the views of the John Templeton Foundation. JMYW acknowledges the financial support from Earth-Life Science Institute and Japan Student Services Organization for postgraduate studies. The source codes for the models used in this study are archived at the Earth-Life Science Institute of the Tokyo Institute of Technology. The data, input and output files necessary to reproduce the figures are available from the authors upon request.References

Abe, Y., 1993. Physical state of the very early Earth. Lithos 30, 223-235.

Agee, C.B., Wilson, N.V., McCubbin, F.M., Ziegler, K., Polyak, V.J., Sharp, Z.D., Asmerom, Y., Nunn, M.H., Shaheen, R., Thiemens, M.H., et al., 2013. Unique meteorite from early Amazonian Mars: Water-rich basaltic breccia Northwest Africa 7034. Science 339, 780-785.

Anderson, R.C., Dohm, J.M., Golombek, M.P., Haldemann, A.F., Franklin, B.J., Tanaka, K.L., Lias, J., Peer, B., 2001. Primary centers and secondary concentrations of tectonic activity through time in the western hemisphere of Mars. Journal of Geophysical Research: Planets 106, 20563-20585.

Andrews-Hanna, J.C., Zuber, M.T., Banerdt, W.B., 2008. The borealis basin and the origin of the Martian crustal dichotomy. Nature 453, 1212-1215.

Becker, H., Horan, M., Walker, R., Gao, S., Lorand, J.P., Rudnick, R., 2006. Highly siderophile element composition of the Earths primitive upper mantle: constraints from new data on peridotite massifs and xenoliths. Geochimica et Cosmochimica Acta 70, 4528-4550.

Borg, L.E., Brennecka, G.A., Symes, S.J.K., 2016. Accretion timescale and impact history of Mars deduced from the isotopic systematics of martian meteorites. Geochimica et Cosmochimica Acta 175, 150-167

Bottke, W.F., Walker, R.J., Day, J.M.D., Nesvorny, D., Elkins-Tanton, L., 2010. Stochastic late accretion to Earth, the moon, and Mars. Science 330, 1527-1530.

Bottke, W.F., Andrews-Hanna, J.C., 2017. A post-accretionary lull in large impacts on early Mars. Nat. Geosci. 10, 344–348

Bouvier, L.C., Costa, M.M., Connelly, J.N., Jensen, N.K., Wielandt, D., Storey, M., Nemchin, A.A., Whitehouse, M.J., Snape, J.F., Bellucci, J.J., et al., 2018. Evidence for extremely rapid magma ocean crystallization and crust formation on Mars. Nature 558, 586-589.30